\documentclass[twocolumn,aps,prl]{revtex4}
\usepackage{amsmath}
\usepackage{graphicx}

\begin{document}

\title{Heat Transfer between Weakly Coupled Systems}
\author{ B.N.J. Persson$^{1,2,3}$, A.I. Volokitin$^{2,4}$ and H. Ueba$^{1}$}

\affiliation{$^1$Division of Nanotechnology and New Functional Material Science,
Graduate School of Science and Engineering,
University of Toyama, Toyama, Japan}
\affiliation{$^2$IFF, FZ-J\"ulich, 52425 J\"ulich, Germany, EU}
\affiliation{$^3$www.MultiscaleConsulting.com}
\affiliation{$^4$Samara State Technical University, 443100 Samara, Russia}

\begin{abstract}
We study the heat transfer between weakly coupled systems with flat interface.
We present simple analytical results which can be used to estimate the heat transfer coefficient.
As applications we consider the heat transfer across solid-solid contacts, and between 
a membrane (graphene) and a solid substrate (amorphous ${\rm SiO_2}$). 
For the latter system the calculated value of the heat transfer coefficient is in good agreement 
with the value deduced from experimental data.
\end{abstract}

\maketitle

\pagestyle{empty}


{\bf 1 Introduction}

Almost all surfaces in Nature and Technology have roughness on many different length scales\cite{PSSR}.
When two macroscopic solids are brought into contact, even if the applied force is very small,
e.g., just the weight of the upper solid block, the pressure in the asperity contact regions 
can be very high, usually close to the yield stress of the (plastically) softer solid.
As a result good thermal contact may occur within each microscopic contact region, but owing
to the small area of real contact the (macroscopic) heat transfer coefficient may still be small.
In fact, recent studies have shown that in the case of surfaces with roughness on many different
length scales, the heat transfer is {\it independent}
of the area of real contact\cite{PLV}. We emphasize that this remarkable and counter-intuitive result is only valid when
roughness occur over several decades in length scale.

For nanoscale systems the situation may be very different. 
Often the surfaces are very smooth with typically nanometer
(or less) roughness on micrometer-sized surface areas, 
and because of adhesion the solids often make contact over a large fraction of the
nominal contact area. 
The heat transfer between solids in perfect contact is usually calculated using
the so called acoustic and diffusive mismatch models\cite{Pohl}, where it is assumed that all phonons scatter
elastically at the interface between two materials. In these  models there is no direct reference to
the nature of the solid-solid interaction across the interface, and the models cannot describe
the heat flow between weakly interacting solids.

Here we will discuss the heat transfer across perfectly flat interfaces.
The theory we present is general,
but we will mainly focus on the case when the
interaction between the solids is very weak, e.g., of the Van der Waals type, as for graphene or
carbon nanotubes on many substrates. We present simple 
analytical results which can be used to estimate the heat transfer coefficient.
We study heat transfer for solid-solid and solid-membrane contacts. 
We consider in detail the heat transfer between graphene and amorphous ${\rm SiO_2}$. 
For this system the calculated value of the heat transfer coefficient is in good agreement 
with the value deduced from experimental data.

\begin{figure}[tbp]
\includegraphics[width=0.4\textwidth,angle=0]{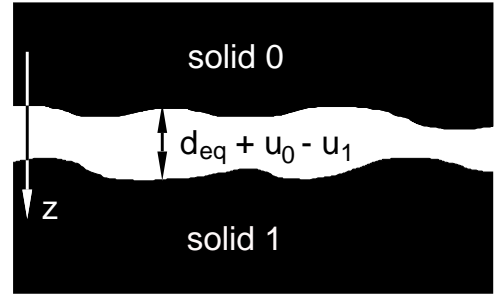}
\caption{
Two solids {\bf 0} and {\bf 1} in contact. The interfacial
surface separation is the sum of the equilibrium separation $d_{\rm eq}$ and the difference 
in the surface displacements $u_0-u_1$, due to thermal movements, 
where both  $u_0$ and  $u_1$  are positive when the displacement point
along the  $z$-axis towards the interior of solid {\bf 1}. Due to interaction between the solids
a perpendicular stress (or pressure) $\pm K(u_0({\bf x},t)-u_1 ({\bf x},t))$ will act on the (interfacial) 
surfaces of the solid.}
\label{picshem}
\end{figure}

\vskip 0.5cm
{\bf 2 Theory}

Consider the interface between two solids, and assume that local thermal equilibrium occurs everywhere except at the 
interface. The energy flow (per unit area) through the interface is given by\cite{PLV}
$$J= \alpha (T_0-T_1),$$
where $T_0$ and $T_1$ are the local temperatures at the interface in solid {\bf 0} and {\bf 1}, respectively.
The stress or pressure acting on the surface of solid ${\bf 1}$ from solid ${\bf 0}$ can be written as
$$\sigma ({\bf x},t) = K[u_0({\bf x},t)-u_1({\bf x},t)],$$
where $u_0$ and $u_1$ are the (perpendicular) surface displacement of solid ${\bf 0}$ and  ${\bf 1}$ 
(see Fig. \ref{picshem}), respectively, and where  $K$
is a spring constant per unit area characterizing the interaction between the two solids.
For weakly interacting solids the parallel interfacial spring constant $K_\parallel $ is usually much smaller than
the perpendicular spring constant  $K_\perp =K$, and we will neglect the heat transfer resulting from 
the tangential interfacial stress associated with thermal vibrations (phonons).

If we define
$$u ({\bf q},\omega)  = {1\over (2\pi )^{3}} \int d^2x dt \ u({\bf x},t) e^{-i({\bf q}\cdot {\bf x} -\omega t)}, $$
we get
$$\sigma ({\bf q},\omega) = K[u_0({\bf q},\omega)-u_1({\bf q},\omega)].\eqno(1)$$
Within linear elasticity theory\cite{PJCP}
$$u_1 ({\bf q},\omega) = M_1({\bf q},\omega) \sigma ({\bf q},\omega), \eqno(2)$$
where  $M_1({\bf q},\omega)$ is determined by the elastic properties of solid ${\bf 1}$.
We consider the heat transfer from solid {\bf 0} to solid {\bf 1}. The displacement of an atom
in solid {\bf 0} is the sum of a contribution derived from the applied stress $-\sigma$, and a stochastic fluctuating contribution
$u_{\rm 0 f}$ due to the thermal movement of the atoms in the solid in the absence of interaction between the solids:
$$u_0 ({\bf q},\omega) = u_{\rm 0 f}({\bf q},\omega)  - M_0({\bf q},\omega) \sigma ({\bf q},\omega), \eqno(3)$$
Combining (1)-(3) gives
$$u_1 ({\bf q},\omega) = {K M_1({\bf q},\omega) \over 1+K [M_0({\bf q},\omega)+ M_1({\bf q},\omega)]} u_{\rm 0 f} ({\bf q},\omega), \eqno(4)$$
$$u_0 ({\bf q},\omega) = {1+K M_1({\bf q},\omega) \over 1+K [M_0({\bf q},\omega)+ M_1({\bf q},\omega)]} u_{\rm 0 f} ({\bf q},\omega). \eqno(5)$$
The energy transferred to solid ${\bf 1}$ from solid ${\bf 0}$ during the time period $t_0$ can be written as
$$\Delta E = \int d^2 x d t \  \dot u_1({\bf x},t) \sigma ({\bf x},t),$$
where  $\dot u = \partial u/ \partial t$. One can also write 
$$\Delta E =(2\pi )^3 \int d^2 q d \omega \  (-i\omega) u_1({\bf q},\omega) \sigma (-{\bf q},-\omega)$$
Using (1), (4) and (5) we obtain
$$\Delta E = (2\pi )^3 \int d^2 q d \omega$$
$$\times  {\omega  K^2 {\rm Im} M_1({\bf q},\omega) \over |1+K [M_0({\bf q},\omega)+ M_1({\bf q},\omega)]|^2} 
\langle |u_{\rm 0 f}({\bf q},\omega)|^2\rangle, \eqno(6)  $$
where we have performed an ensemble (or thermal) average denoted by  $ \langle .. \rangle $. 
Next, note that (see Appendix A)
$$\langle |u_{\rm 0 f}({\bf q},\omega)|^2\rangle =
{A_0 t_0 \over (2\pi )^3} C_{uu}({\bf q},\omega), \eqno(7)$$
where $A_0$ is the surface area, and 
$$C_{uu}({\bf q},\omega) =  {1\over (2\pi )^{3}} \int d^2x dt \ 
\langle u_{\rm 0 f} ({\bf x},t) u_{\rm 0 f} (0,0)\rangle e^{i({\bf q}\cdot{\bf x}-\omega t)}, $$
is the displacement correlation function.
Using the fluctuation-dissipation theorem\cite{fluctuation} we have (see also Appendix B and C)
$$C_{uu}({\bf q},\omega)= {2\over (2\pi)^3} {\Pi (\omega) \over \omega} {\rm Im}  M_0 ({\bf q}, \omega)\eqno(8)$$
where $\Pi (\omega)= \hbar \omega \left [{\rm exp} (\hbar \omega / k_{\rm B} T_0) - 1\right ]^{-1}.$
Substituting (7) in (6) and using (8) gives the heat current $J_{0\rightarrow 1} = \Delta E/ A_0 t_0$ from solid  ${\bf 0}$ to  solid ${\bf 1}$:
$$J_{0\rightarrow 1} 
= {4\over (2 \pi)^3} \int d^2 q \int_0^\infty d \omega \ \Pi (\omega)$$
$$\times { {\rm Im} K M_0({\bf q},\omega ) {\rm Im} K M_1({\bf q},\omega )
\over | 1+K[M_0({\bf q},\omega)+M_1({\bf q},\omega)]|^2}, $$
A similar equation with $T_0$ replaced by $T_1$ gives the energy transfer from solid ${\bf 1}$ to solid ${\bf 0}$, and the
net energy flow $J=J_{0\rightarrow 1}-J_{1\rightarrow 0}$.
The heat transfer coefficient $\alpha = (J_{0\rightarrow 1}-J_{1\rightarrow 0})/(T_0-T_1)$ gives in the limit
$(T_0-T_1)\rightarrow 0$:
$$\alpha 
= {4\over (2 \pi)^3} \int d^2 q \int_0^\infty d \omega \ {\partial \Pi (\omega)\over \partial T}$$
$$\times { {\rm Im} K M_0({\bf q},\omega ) {\rm Im} K M_1({\bf q},\omega )
\over | 1+K[M_0({\bf q},\omega)+M_1({\bf q},\omega)]|^2}, 
\eqno(9)$$

To proceed we need expressions for $ M_0({\bf q},\omega) $ and $ M_1({\bf q},\omega) $. 
Here we give the $M$-function for (a) solids, (b) liquids and (c) membranes.

\vskip 0.3 cm
{\bf (a) Solids}

For an elastic solid we have\cite{PJCP,Ryberg}
$$M= {i \over \rho c^2_{\rm T}} {p_{\rm L}(q,\omega) \over S(q,\omega)} 
\left ({\omega \over c_{\rm T}}\right )^2\eqno(10)$$
where
$$S= \left [\left ({\omega \over c_{\rm T}}\right )^2-2q^2\right]^2+4q^2p_{\rm T}p_{\rm L} $$
$$p_{\rm L} = \left [\left ({\omega \over c_{\rm L}}\right )^2 -q^2+i0\right ]^{1/2} $$
$$ p_{\rm T} = \left [\left ({\omega \over c_{\rm T}}\right )^2 -q^2+i0\right ]^{1/2} $$ 
where $c_{\rm L}$, $c_{\rm T}$ and $\rho$ are the longitudinal and transverse 
sound velocities, and the mass density, respectively.

\vskip 0.3 cm
{\bf (b) Liquids}

This case can be obtained directly from the solid case by letting $c_{\rm T}\rightarrow 0$:
$$M = {i p_{\rm L} \over \rho \omega^2} = {i \over \rho \omega^2} 
\left [\left ({\omega \over c_{\rm L}}\right )^2 -q^2+i0\right ]^{1/2}\eqno(11)$$

\vskip 0.3 cm
{\bf (c) Membranes}

We assume that the out-of-plane displacement $u({\bf x},t)$ satisfies
$$ \rho_0 {\partial^2 u \over \partial t^2} = -\kappa \nabla^2 \nabla^2 u  +\sigma, \eqno(12)$$
where $\rho_0 = n_0 m_0$ is the mass density per unit {\it area} of the 2D-system ($m_0$ is the atom mass and $n_0$ the number of
atoms per unit area), $\kappa $ is the bending elasticity (for graphene,
 $\kappa \approx 1 \ {\rm eV} $ \cite{Fasolino}), and $\sigma({\bf x},t)$
an external stress acting perpendicular to the membrane (or $xy$-plane).
Using the definition  $ M({\bf q},\omega) = u({\bf q},\omega)/\sigma({\bf q},\omega)$  from (12) we get
$$M = {1\over \kappa q^4 -\rho_0 \omega^2 - i 0^+}.\eqno(13)$$

\vskip 0.5 cm
{\bf 3 Some limiting cases}

Assuming weak coupling between the solids (i.e., $K$ is small) and high enough temperature, (9) reduces to
$$\alpha
= {4K^2 \over (2 \pi)^3} \int d^2 q \int_0^\infty d \omega \ {\partial \Pi (\omega)\over \partial T}
{\rm Im} M_1({\bf q},\omega) {\rm Im} M_0({\bf q},\omega). \eqno(14)$$
In the opposite limit of strong coupling ($K\rightarrow \infty$) we get
$$\alpha
= {4\over (2 \pi)^3} \int d^2 q \int_0^\infty d \omega \ {\partial \Pi (\omega)\over \partial T} 
{ {\rm Im}  M_0({\bf q},\omega) {\rm Im}  M_1({\bf q},\omega)
\over | M_0({\bf q},\omega)+M_1({\bf q},\omega)|^2 },\eqno(15)$$
which does not depend on $K$. Note also that for very low temperatures only very low frequency phonons
will be thermally excited. Assuming a semi-infinite solid,  
as $\omega \sim q \rightarrow 0$, from (10) we have
$|M|\approx 1 /(\rho c \omega) \rightarrow \infty $ (where $c$ is the sound velocity and 
$\rho$ the mass density). Thus, at low enough temperature 
(9) reduces to (15) i.e., for very low temperatures the heat transfer is {\it independent} of the 
strength of the interaction across the
interface. The physical reason for this is that at very low temperature the wavelength of the phonons becomes very long and
the interfacial interaction becomes irrelevant. The transition between the two regions of behavior occurs when
$K |M| \approx 1$. Since $|M| \approx 1/ (\rho c \omega)$ we get $K \approx \rho c \omega$.
But $\hbar \omega \approx k_{\rm B} T$ and defining the thermal length $\lambda_{\rm T} = c/\omega = c \hbar / k_{\rm B} T$
we get the condition $K \approx \rho c^2 /\lambda_{\rm T}$. Since the elastic modulus $E\approx \rho c^2$ we get
$K \approx E/\lambda_{\rm T}$. We can define a spring constant between the atoms in the solid via
$k'=Ea$, where $a$ is the lattice constant. Since $K = k/a^2$ we get $k \approx  (a/\lambda_{\rm T}) k' $ as the condition 
for the transition between the two different regimes in the heat transfer behavior. For most solids at room temperature
$\lambda_{\rm T} \approx a$, but at very low temperatures $\lambda_{\rm T}>> a$ which means that even a very weak (soft) interface
(for which  $k$ is small), will appear as very strong (stiff) with respect to the heat transfer at low temperatures.

Let us consider the case where the two solids are identical, and assume strong coupling where (15) holds. For this case we do not 
expect that the interface will restrict the energy flow. If we consider high temperature the kinetic energy per atom in solid {\bf 0}
will be $\sim k_{\rm B} T_0$ so the energy density  $Q \approx k_{\rm B} T_0 / a_0^3$ (where $a_0$ is the lattice constant). Thus if solid
{\bf 1} is at zero temperature we expect the energy flow current across the interface to be of order $J\approx Q c/4$ (the
factor of $1/4$ results from the fact that only half the phonons propagate in the positive $z$-direction and the
average velocity of these phonons in the $z$-direction is $c/2$). Thus we expect
$\alpha \approx k_{\rm B} c /(4a_0^3) $. This result follows also from (15) if we notice that for  $M_0 =M_1 $ and high temperatures
$$\alpha
= {k_{\rm B}  \over (2 \pi)^3} \int d^2 q \int_0^\infty d \omega \  \left [ {{\rm Im} M_0({\bf q},\omega) \over 
 |M_0({\bf q},\omega)|} \right ]^2 \eqno(16)$$
If we assume for simplicity that $M_0$ is given by (11) (but the same qualitative result is obtained for solids) 
then the factor involving $M_0$ is equal to unity for $\omega > c_{\rm L} q$ and zero otherwise. Thus (16) reduces to
$$\alpha
=  {  k_{\rm B}  \over (2 \pi)^2}  \int_0^{c_{\rm L} q_{\rm c}}  d \omega  \int_0^{\omega/ c_{\rm L}} d q \ q
=  {\pi c_{\rm L}  k_{\rm B} \over 24 a_0^3}$$
where we have used that $q_{\rm c} \approx \pi /a_0$. Thus for identical materials and strong coupling (9) reduces
to the expected result.

Let us now briefly discuss the temperature dependence of the heat transfer coefficient for high and low temperatures.
For very low temperatures $\alpha $ is given by (15). Consider first a solid in contact with a solid or liquid. For these
cases it follows that $ M \sim 1/ \omega$ (where we have used that $ \omega \sim q$) so the temperature dependence 
of the heat transfer coefficient is determined by the term
$$\alpha
\sim  \int_0^\infty d \omega \ {\partial \Pi (\omega)\over \partial T} \omega^2$$
where we also have used that $d^2 q \sim \omega^2$. Thus we get
$$\alpha
\sim  \int_0^\infty d \omega { {\rm exp}(\hbar \omega /k_{\rm B} T) \over  \left [{\rm exp}(\hbar \omega /k_{\rm B} T)-1 \right ]^{2}}
\left ({\hbar \omega \over k_{\rm B} T} \right )^2 \omega^2 \sim T^3$$
For high temperatures, and assuming weak coupling, one obtains in the same way from (14) that $\alpha $ is 
temperature independent. However, the spring constant (per unit area) $K $ may depend on the temperature, e.g., as a result
of thermally induced rearrangement of the atoms at the contacting interface or thermally induced increase in the separation
of the two surfaces at the interface which may be particularly important for weakly interacting systems.
The temperature dependence of $\alpha $ for the case of solid-solid and solid-membrane contacts will be discussed in Sec.5.

\vskip 0.5 cm
{\bf 4 Phonon heat transfer at disordered interfaces: friction model}

At high temperature and for atomically disordered interfaces, the interfacial atoms will 
perform very irregular, stochastic motion. In this case
the heat transfer coefficient $\alpha$ can be obtained (approximately) from a classical
``friction'' model. 
This treatment does not take into account in a detailed way the restrictions on the energy transfer process 
by the conservation of parallel momentum, which arises for periodic (or homogeneous) solids. See also Appendix D.

Let us assume that solid {\bf 0} has a lower maximal phonon frequency than solid {\bf 1}.
In this case, most elastic waves (phonons) in solid  {\bf 0} can in principle propagate into solid  {\bf 1},
while the opposite is not true, since a phonon in solid  {\bf 1} with higher 
energy than the maximum phonon-energy in solid  {\bf 0} will, because of energy conservation,
be totally reflected at the interface between the solids.

Consider an atom in solid {\bf 0} (with mass $m_0$) 
vibrating with the velocity $({\bf v}_\parallel, v_\perp)$.
The atom will exert a fluctuating force on solid {\bf 1} which will result in elastic waves (phonon's) being
excited in solid {\bf 1}. The emitted waves give rise to a friction force acting on 
the atom in solid {\bf 0} (from solid {\bf 1}), which we can write as\cite{Ryberg}
$${\bf F}_f = -m_0 \eta_\parallel {\bf v}_\parallel -m_0 \eta_\perp {\bf v}_\perp , $$
and the power transfer to solid {\bf 1} will be
$$P=-\langle {\bf F}_f\cdot {\bf v} \rangle = m_0 \eta_\parallel \langle v^2_\parallel \rangle +  m_0 \eta_\perp  \langle v^2_\perp \rangle .  $$
At high temperatures
$$ m_0 \langle v^2_\parallel \rangle = 2 k_{\rm B}T_0, \ \ \ \ \ \ 
m_0 \langle v^2_\perp \rangle = k_{\rm B}T_0. $$
Hence
$$P=(2 \eta_\parallel + \eta_\perp ) k_{\rm B}T_0. $$
A similar formula (with $T_0$ replaced by $T_1$) gives the power transfer from solid {\bf 1} to solid {\bf 0}. Hence
$$J = n_0 (2 \eta_\parallel + \eta_\perp ) k_{\rm B} (T_0-T_1),$$  
where $n_0=1/a_0^2$ is the number of interfacial atoms per unit area in solid {\bf 0}. Thus we get
$$\alpha = n_0 (2 \eta_\parallel + \eta_\perp ) k_{\rm B}.\eqno(17)$$  
For weak interfacial coupling we expect $\eta_\perp >> \eta_\parallel $, and we can neglect the $\eta_\parallel $-term in (17).

The damping or friction coefficient  $\eta_\perp$  due to phonon emission 
was calculated within elastic continuum mechanics in Ref. \cite{Ryberg}. We have
$$\eta_{\perp} \approx {k^2 \xi' \over \rho_1 m_0 c_{\rm T}^3} =   {K^2 a_0^4 \xi' \over \rho_1 m_0 c_{\rm T}^3} , \eqno(18)$$
where $\xi' \approx 0.13$ (see Ref. \cite{explain}).
Using that $n_0 = 1/a_0^2$ and substituting (18) in (17) gives
$$\alpha = {k_{\rm B} K^2 \xi' \over  \rho_0 \rho_1 c_{\rm T}^3},\eqno(19)$$
where
$$\xi'= {1\over 8 \pi} {\rm Re} \int_0^\infty 
dx  {2 \left (\gamma-x\right )^{1/2} 
\over \left (1-2x\right  )^2 +4 \left (1-x\right )^{1/2} \left (\gamma-x\right )^{1/2}}$$
where $\gamma = (c_{\rm T}/c_{\rm L})^2$, and 
where $\rho_0 = n_0 m_0$ is the  (one atomic layer) mass per unit surface area of solid {\bf 0}. 
There are two contributions to the integral $\xi'$. One is derived from the region $x<1$ where the integral clearly
has a non-vanishing real part. This contribution correspond to excitation of transverse and longitudinal acoustic phonons.
The second contribution arises from the vicinity of the point (for $x>1$) where the denominator vanish.
This pole contribution correspond to excitation of surface (Rayleigh) waves.  
As shown in Ref. \cite{Ryberg}, about $65\%$ of the radiated energy is due to the surface (Rayleigh) phonons, and the rest by
bulk acoustic phonons.

We emphasize that (19) is only valid for high temperatures and weak coupling. A more general
equation for the heat transfer between solids when the phonon emission occur incoherently is derived
in Appendix D:
$$\alpha 
\approx {4 A^* \over (2 \pi)^3} \int_0^\infty d \omega \ {\partial \Pi (\omega)\over \partial T} 
{ {\rm Im} K M_0(\omega ) {\rm Im} K M_1(\omega )
\over | 1+K[ M_0(\omega)+M_1(\omega)]|^2},\eqno(20)$$
where
$$M(\omega) =  {1\over A^*} \int_{q<q_{\rm c}} d^2q \ M({\bf q},\omega),$$
where the integral is over $|{\bf q}|<q_{\rm c}$, where 
$\pi q_{\rm c}^2 = A^*$. The cut-off wavevector  $q_{\rm c} $ is the smallest of  $q_1$ and $q_2$, where
$\pi q_1^2 = (2\pi )^2/a_0^2$ (where  $a_0 $ is the lattice 
constant) and where $q_2=k_{\rm B}T/\hbar c_0 $ (where  $c_0 $ is the smallest sound velocity of solid {\bf 0}) is
the thermal wavevector. For high temperatures and weak coupling, for an Einstein model of solid
{\bf 0},  (20) reduces to (19) (see Appendix E).

\vskip 0.5 cm
{\bf 5 Numerical results}

We now present some numerical results to illustrate the theory presented above. We consider (a) solid-solid, 
(b) solid-liquid and (c) solid-membrane systems.

\vskip 0.3 cm
{\bf (a) Solid-Solid}

\begin{figure}[tbp]
\includegraphics[width=0.4\textwidth,angle=0]{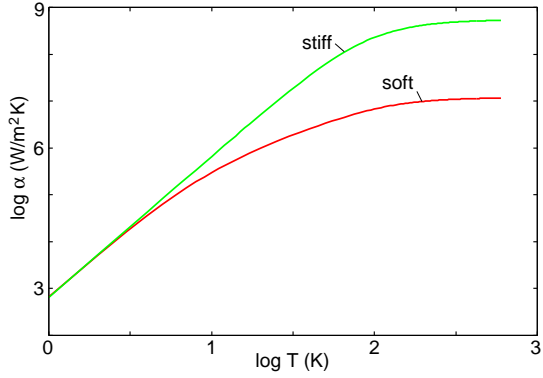}
\caption{
The logarithm (with 10 as basis) of the heat transfer coefficient as a function of the 
logarithm of the temperature for weakly interacting solids (soft) 
with $K= 2.52\times 10^{19} \ {\rm N/m^3}$, and for solids which interact stronger 
(stiff) with 10 times larger $K$. The dotted line has the slope $3$ corresponding the a $\sim T^3$
temperature dependence.}
\label{new.all.solid.solid.stiff.soft}
\end{figure}

\begin{figure}[tbp]
\includegraphics[width=0.4\textwidth,angle=0]{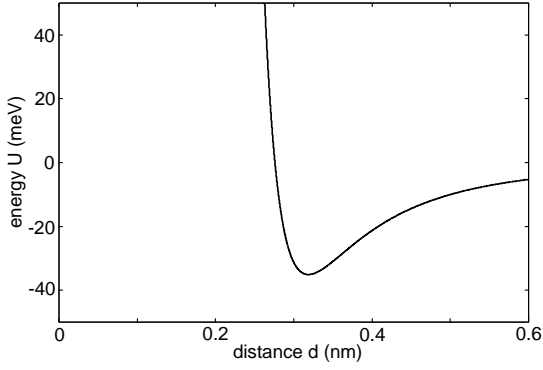}
\caption{
The calculated ${\rm graphene}-{\rm a}$-${\rm SiO_2}$ interaction energy $U(d)$ per graphene carbon atom, as a function
of the separation $d$ (in nm) between the center of a graphene carbon atom and the center of the first layer
of substrate atoms. See text for details.}
\label{d.nm.U.meV.per.C.atom.graphene.SiO2}
\end{figure}

We consider the heat transfer between two solids with perfectly flat contacting surfaces. We take the
sound velocities and the mass density and the (average) lattice constant to be that of ${\rm SiO_2}$.
We consider two cases: weakly interacting solids (soft interface) with $K= 2.52\times 10^{19} \ {\rm N/m^3}$ (see Fig. 
\ref{d.nm.U.meV.per.C.atom.graphene.SiO2}), and 
solids with stronger interaction (stiff interface), with 10 times larger $K$. In Fig. \ref{new.all.solid.solid.stiff.soft}
we show the heat transfer coefficient as a function of temperature. Note that
for high temperatures $\alpha $ is nearly 100 times larger for the stiff case as compared to the soft case. This result is expected based on
Eq. (9b) which shows that $\alpha \sim K^2$ as long as $K$ is not too large or the temperature too low.
For low temperatures both cases gives very similar results, and for $T < 3 \ {\rm K}$ the heat transfer coefficient $\alpha \sim T^3$.
The reason for why at low temperatures the heat transfer is independent of the strength of the interfacial interaction was explained
in Sec. 3, and is due to the long wavelength of the thermally excited  phonons at low temperature.

\begin{figure}[tbp]
\includegraphics[width=0.4\textwidth,angle=0]{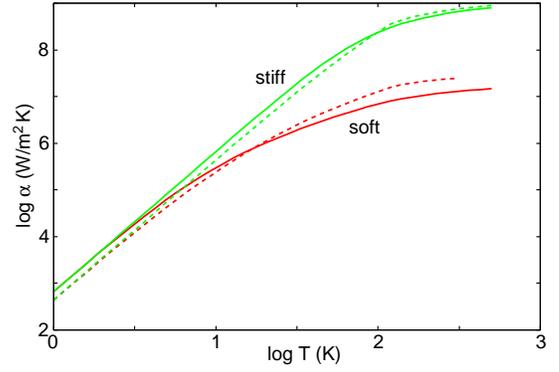}
\caption{
The logarithm (with 10 as basis) of the heat transfer coefficient as a function of the 
logarithm of the temperature for weakly interacting solids (soft) 
with $K= 2.52\times 10^{19} \ {\rm N/m^3}$, and for solids which interact stronger 
(stiff) with 10 times larger $K$. The solid lines are for coherent phonon 
transmission (from Fig. \ref{new.all.solid.solid.stiff.soft}),
and the dashed lines for incoherent phonon transmission.}
\label{hard.soft.coherent.incoherent}
\end{figure}

For incoherent phonon transmission, using (20) we obtain the result shown by dashed curves in 
Fig. \ref{hard.soft.coherent.incoherent}. For both the soft and stiff interface the results obtained assuming coherent and incoherent phonon
transmission are similar.

\vskip 0.3 cm
{\bf (b) Solid-Liquid}

\begin{figure}[tbp]
\includegraphics[width=0.4\textwidth,angle=0]{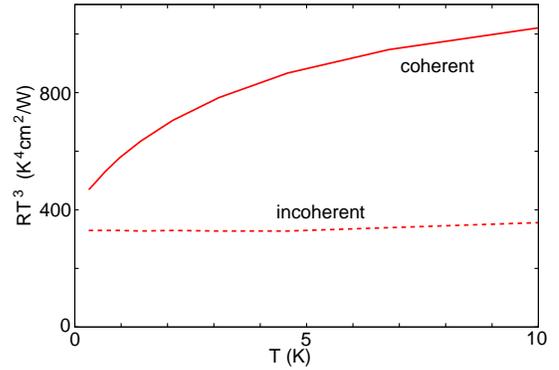}
\caption{
The calculated contact resistance (multiplied by  $T^3$) between liquid $^4{\rm He}$
and {\rm a}-${\rm SiO_2}$ as a function of the temperature, with $K= 1.18\times 10^{19} \ {\rm N/m^3}$.}
\label{all.He}
\end{figure}

Heat transfer between liquid $^4{\rm He}$ and solids was studied by Kapitza\cite{Kapitza} $\sim 60$ years ago,
and $R=  1/\alpha $ is usually denoted as the Kapitza resistance\cite{Pohl,Pollack}.
Let us apply the theory to the heat transfer between liquid $^4{\rm He}$
and {\rm a}-${\rm SiO_2}$. In the calculation we use a He-substrate potential
with well depth $10.2 \ {\rm meV}$ and $^4{\rm He}$-substrate equilibrium bond distance
$d_{\rm eq} = 2.2 \ {\rm \AA}$ which agree with the model parameters used in Ref. \cite{Massimo}.
With these parameters we get the perpendicular He-substrate vibration frequency
$\omega_\perp \approx 91 \ {\rm cm}^{-1}$ and the spring constant $K= 1.18\times 10^{19} \ {\rm N/m^3}$.
In Fig. \ref{all.He} we show the  calculated contact resistance $R$ (multiplied by $T^3$), as a function of the temperature $T$.
In this calculation we have assumed that all the parameters (e.g., $^4{\rm He}$ sound velocity $c_0$ and mass density $\rho_0$) 
characterizing the system are temperature independent\cite{addpara}.
The Kapitza resistance has been measured (for $T > 1 \ {\rm K}$) 
for liquid $^4{\rm He}$ in contact with Quartz\cite{Challis} and Sapphire\cite{Gittleman}
and scales roughly with temperature as $T^{-3}$,
and the magnitude for $T=1 \ {\rm K}$ is roughly 10 times smaller than
our calculated result assuming incoherent phonon transfer.
Experiment show that for $ T < 0.5 \ {\rm K}$ the Kapitza resistance
increase much faster with decreasing temperature than expected from the $R\sim T^{-3}$-dependence 
predicted by our theory and most other theories. It is not clear what may be the origin of this discrepancy,
but it has been suggested to be associated with surface roughness. 
Unfortunately, most measurements of the Kapitza resistance was performed before
recent advances in Surface Science, and many of the studied systems are likely to have oxide and 
unknown contamination layers, which may explain the large fluctuations in the measured contact resistance
for nominally identical systems.

\vskip 0.3 cm
{\bf (c) Solid-Membrane}

From (13) we get:
$${\rm Im} M_0({\bf q},\omega) = \pi \delta (\kappa q^4 -\rho_0 \omega^2) = {\pi \over 2 \rho_0 \omega_1} \delta (\omega - \omega_1)\eqno(21)$$
where $\omega_1 = (\kappa /\rho_0)^{1/2} q^2 = c(q) q$, where we have defined the velocity $ c(q)= (\kappa /\rho_0)^{1/2} q$.  
Substituting (21) in (9) and assuming high temperatures so that $\Pi (\omega) \approx k_{\rm B} T_0 $ gives:
$$J_{0\rightarrow 1}  
 =  {k_{\rm B} T_0\over 2 \pi \rho_0}  \int_0^\infty dq \  {q\over \omega_1}  K^2 {\rm Im} M_1({\bf q},\omega_1) . $$
The heat transfer coefficient $\alpha = (J_{0\rightarrow 1}-J_{1\rightarrow 0})/(T_0-T_1)$ is given by
$$ \alpha =  {k_{\rm B} \over 2 \pi \rho_0}  \int_0^\infty dq \  {q\over \omega_1}  K^2 {\rm Im} M_1({\bf q},\omega_1) .\eqno(22)$$
Using the expression for $M_1({\bf q}, \omega)$ derived in \cite{PJCP,Ryberg} and $\omega_1 = c(q) q$ gives
$$ \alpha =  {k_{\rm B} K^2 \xi \over  \rho_0 \rho_1 c_{\rm T}^3}, \eqno(23)$$
where
$$ \xi =  {1 \over 2 \pi}  \int_0^{q_{\rm c}} dq \  {1\over q} {c(q)\over c_{\rm T}} $$ 
$$ \times {\rm Re} \left( \left [{c^2(q)\over c_{\rm L}^2}-1\right ]^{1/2} \over 
\left [{c^2(q)\over c_{\rm T}^2}-2\right ]^2 +4\left [{c^2(q)\over c_{\rm T}^2}-1 \right ]^{1/2} 
\left [{c^2(q)\over c_{\rm L}^2}-1\right ]^{1/2} \right ), $$
where $c_{\rm L}$, $c_{\rm T}$ and $\rho_1$ are the longitudinal and transverse 
sound velocities, and the mass density, respectively, of solid ${\bf 1}$.
The cut off wavevector $q_{\rm c} \approx \pi /a_1$ ($a_1$ is the lattice constant, or the average distance between two 
nearby atoms) of solid {\bf 1}.

There are two contributions to the integral $\xi$. One is derived from $c(q) > c_{\rm L}$,
but for graphene on a-${\rm SiO_2}$, this gives only $\sim 10\%$ of the contribution to the integral.
For $c(q) < c_{\rm L}$ the term after the Re operator is purely imaginary (and will therefore not contribute
to the integral), {\it except} for the case where the denominator vanish. It is found that
this pole-contribution gives the main contribution ($\sim 90\%$) to the integral, and corresponds to the excitation of
a Rayleigh surface (acoustic) phonon of solid {\bf 1}. This process involves 
energy exchange between a bending vibrational mode 
of the graphene and a Rayleigh surface phonon mode of solid {\bf 1}. The denominator vanish when $c(q)=c_{\rm R}$ where
$$\left [{c^2_{\rm R}\over c_{\rm T}^2}-2\right ]^2 - 4\left [1-{c^2_{\rm R}\over c_{\rm T}^2}\right ]^{1/2} 
\left [1-{c^2_{\rm R}\over c_{\rm L}^2}\right ]^{1/2} =0 . $$
Note that the 
Rayleigh velocity $c_{\rm R} < c_{\rm T}$ but close to $c_{\rm T}$. 
For example, when $c_{\rm L}/ c_{\rm T} = 2$,  $c_{\rm R} \approx 0.93 c_{\rm T}$,
and the pole contribution to the integral in $\xi$ is $0.083$.
Note that (23) is of the same form as (19), and since $\xi'  \approx 0.13 \approx \xi $ they give very similar results. 

In the model above the heat transfer between the solids involves a single bending mode of the
membrane or 2D-system. In reality there will always be some roughness at the interface which will 
blurred the wavevector conservation rule. We therefore expect a narrow band of bending modes to be involved in
the energy transfer, rather than a single mode. Nevertheless, the model study above 
assumes implicitly that, due to lattice non-linearity (and defects), there exist phonon scattering 
processes which rapidly transfer energy to the bending mode
involved in the heat exchange with the substrate. 
This requires very weak coupling to the substrate, so that the
energy transfer to the substrate is so slow that the bending mode can be re-populated by phonon scattering 
processes in the 2D-system, e.g., from the in-plane phonon modes,
in such a way that its population is always close to what would be the case if complete thermal equilibrium 
occurs in the 2D-system. This may require high temperature in order for multi-phonon
scattering processes to occur with enough rates. 

We now consider graphene on amorphous ${\rm SiO_2}$. Graphene, the
recently isolated 2D-carbon material with unique properties due to its linear electronic
dispersion, is being actively explored for electronic applications\cite{Geim}. 
Important properties are the high mobilities reported especially in suspended graphene, the fact that graphene is the ultimately thin
material, the stability of the carbon-carbon bond in graphene, the ability to induce a band gap by electron confinement in graphene
nanoribbons, and its planar nature, which allows established pattering and etching techniques to be applied.
Recently it has been found that the heat generation in graphene field-effect transistors can result 
in high temperature and
device failure\cite{Freitag}. Thus, it is important to understand the the mechanisms which influence the heat flow. 

The ${\rm graphene}-{\rm a}$-${\rm SiO_2}$ interaction is probably of the Van der Waals type. In Ref. \cite{Pop} the interaction between
the graphene C-atoms and the substrate Si and O atoms was assumed to be described by Lennard-Jones (LJ) pair-potentials. 
Here we use a simplified picture where the substrate atoms form a simple cubic lattice with the lattice constant
determined by $a_1=(\bar m/\rho_1)^{1/3} \approx 0.25 \ {\rm nm}$, 
where $\bar m = (m_{\rm Si}+2 m_{\rm O})/3 \approx 3.32\times 10^{-26} \ {\rm kg}$ is the average substrate atomic mass,
and $\rho_1\approx 2200 \ {\rm kg/m^3} $ the mass density of a-${\rm SiO}_2$. We also use the effective LJ energy
parameter, $\epsilon = (\epsilon_{\rm Si}+2\epsilon_{\rm O})/3\approx 5.3 \ {\rm meV}$, and the bond-length parameter 
$\sigma = (\sigma_{\rm Si}+2\sigma_{\rm O})/3 \approx 0.31 \ {\rm nm}$. With these parameters we can calculate 
the ${\rm graphene}-$a-${\rm SiO_2}$ interaction energy, $U(d)$, per graphene carbon atom, as a function
of the separation $d$ between the center of a graphene carbon atom and the center of the first layer
of substrate atoms. We find (see Fig. \ref{d.nm.U.meV.per.C.atom.graphene.SiO2}) the
${\rm graphene}-$a-${\rm SiO_2}$ binding energy $E_{\rm b} = -U(d_{\rm eq}) \approx \ 35 \ {\rm meV}$ 
per carbon atom, and the force constant $k=U''(d_{\rm eq})$
(where $d_{\rm eq}\approx 0.32 \ {\rm nm}$ is the equilibrium separation)
$k= K a_0^2 = 2.4 \ {\rm N/m}$ per carbon atom. This gives the perpendicular ${\rm graphene}-{\rm a}$-${\rm SiO_2}$ (uniform) vibration
frequency $\omega_\perp \approx (k/m_0)^{1/2} \approx \ 55 \ {\rm cm}^{-1}$, 
which is similar to what is observed for the perpendicular 
vibrations of linear alkane molecules on many surfaces (e.g., about $ 50-60 \ {\rm cm}^{-1}$ for alkanes on
metals and on hydrogen terminated diamond C(111)\cite{Woell}). 
Using $K = k/ a_0^2 = 1.82\times 10^{20} \ {\rm N/m^3}$, and the transverse and longitudinal sound velocities of
solid {\bf 1} ($c_{\rm T} = 3743 \ {\rm m/s}$ and $c_{\rm L} = 5953 \ {\rm m/s}$), from (19) we obtain 
$\alpha \approx 3\times 10^8 \ {\rm W/Km^2}$. 

The heat transfer coefficient between graphene and a {\it perfectly flat} a-${\rm SiO_2}$ substrate has not been measured
directly, but measurements of the heat transfer between carbon nanotubes and sapphire 
by Maune et al\cite{Maune} indicate that it may be of order $\alpha \approx 8\times 10^8  \ {\rm W/m^2K}$.
This value was deduced indirectly by measuring the breakdown voltage of carbon nanotubes, which could be related to the
temperature increase in the nanotubes. Molecular dynamics calculations\cite{Pop} for nanotubes on a-${\rm SiO_2}$  gives 
$\alpha \approx 3 \times 10^8  \ {\rm W/m^2K} $ (here it has been assumed that the contact width between the nanotube and
the substrate is $1/5$ of the diameter of the nanotube). 
Finally, using a so called 3 $\omega $ method, Chen et al\cite{Chen} have measured the heat 
transfer coefficient $\alpha \approx 2 \times 10^8 \ {\rm W/m^2 K}$.

\begin{figure}[tbp]
\includegraphics[width=0.4\textwidth,angle=0]{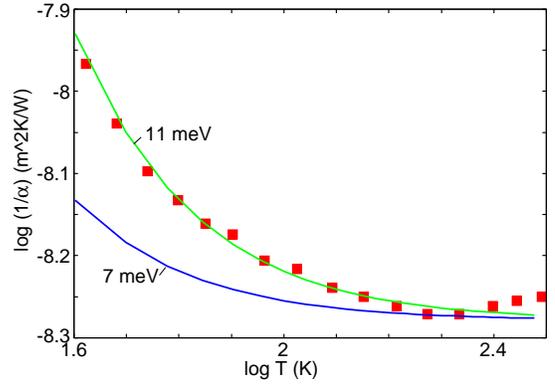}
\caption{
The logarithm of the contact resistance for graphene on {\rm a}-${\rm SiO_2}$ 
as a function of the logarithm of the temperature.
Square symbols: measured data from Ref. \cite{Chen}. Solid lines: the calculated contact resistance using
Eq. (24) with $\hbar \omega_0 = 11 \ {\rm meV}$ (upper curve) and  $\hbar \omega_0 = 7 \ {\rm meV}$ (lower curve).}
\label{graphene.Temp.R.using.alpha=1.9E+08}
\end{figure}

\begin{figure}[tbp]
\includegraphics[width=0.4\textwidth,angle=0]{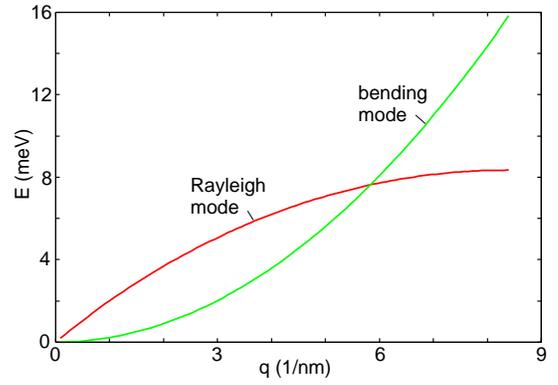}
\caption{
The frequency of the graphene bending mode becomes equal to the frequency of the Rayleigh mode
when $\omega_{\rm bend} (q) = \omega_{\rm R}(q)$. 
The Rayleigh mode dispersion was measured for  $\alpha  $-quartz (0001)\cite{Steurer} and the bending
mode dispersion was calculated using  $\omega_{\rm bend} = (\kappa /\rho_0)^{1/2}q^2 $ with the bending stiffness
 $\kappa = 1.1 \ {\rm eV}  $ as obtained in Ref. \cite{Fasolino}.
}
\label{q.phononR.dispersion}
\end{figure}

We now discuss the temperature dependence of the heat transfer coefficient. If we assume that most of the heat transfer is via 
a substrate phonon mode at the frequency $\omega_0$, then the temperature dependence of $\alpha$ should be given by
$${d \Pi (\omega_0) \over d T} ={x^2  e^x \over (e^x -1)^2},  \eqno(24)$$
where $x=\hbar \omega_0 /k_{\rm B}T$. 
In Fig. \ref{graphene.Temp.R.using.alpha=1.9E+08} we show  the temperature dependence of the heat transfer coefficient 
measured by Chen et al  \cite{Chen} for $42 \ {\rm K} < T < 310 \ {\rm K}$. The solid lines has been
calculated using (24) with $\hbar \omega_0
= 11 \ {\rm meV}$ (upper curve) and $7 \ {\rm meV}$ (lower curve). 
In our model all the vibrational modes have linear dispersion relation, e.g., $\hbar \omega = c_{\rm R} q $
for the Rayleigh mode, and the frequency where the graphene bending mode become equal to the frequency of the Rayleigh mode
will occur at higher frequency than expected using the measured Rayleigh mode dispersion relation. 
This is illustrated in Fig. \ref{q.phononR.dispersion} where we show the 
measured Rayleigh mode dispersion for  $\alpha  $-quartz (0001)\cite{Steurer}. Note that the frequencies of the bending mode and
the Rayleigh mode become equal when $\hbar \omega \approx 7 \ {\rm meV}$. However, using this excitation energy in
(24) gives too weak temperature dependence. There are two possible explanations for this: 

(a) In an improved calculation using the measured
dispersion relations for the substrate phonon modes, emission of bulk phonons may become more important than in the present study
where we assumed the linear phonon-dispersion is valid for all wavevectors $q$. This would make higher excitation energies more important
and could lead to the effective (or average) excitation energy $11 \ {\rm meV}$ necessary to fit the observed temperature dependence.

(b) As pointed out in Sec. 3, the model developed 
above for the heat transfer involves a single, or a narrow band, of bending modes of the
membrane or 2D-system. In order for this model to be valid, the coupling to the substrate must be so weak that the
energy transfer to the substrate from the bending mode occurs so slowly that the mode can be re-populated by phonon scattering 
processes, in such a way that its population is always close to what is expected if full thermal equilibrium 
would occur within the 2D-system. This may require high temperature in order for multi-phonon
scattering processes to occur by high enough rates. 
This may contribute to the decrease in the heat transfer coefficient observed for the  ${\rm graphene}-{\rm a}$-${\rm SiO_2}$ system 
below room temperature\cite{Chen}.

Finally, we note that recently it has been suggested\cite{Rotkin,Freitag} that the heat transfer between graphene and a-${\rm SiO_2}$ may involve
photon tunneling\cite{rev1}. That is, coupling via the electromagnetic field between electron-hole pair
excitations in graphene and optical phonons in a-${\rm SiO_2}$. However, our calculations indicate that
for graphene adsorbed on  a-${\rm SiO_2}$ the field coupling gives a negligible contribution to the heat transfer\cite{PUeba}.

\begin{figure}[tbp]
\includegraphics[width=0.5\textwidth,angle=0]{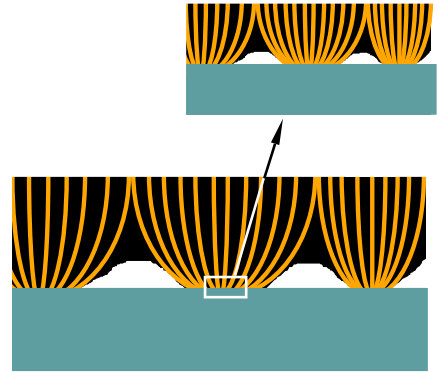}
\caption{
Heat flow in the contact region between a rigid block with a flat surface (bottom) and en elastic
solid with a randomly rough surface (top). The orange lines denote the heat current flux lines
in the upper solid. The heat current filaments expand laterally until the filaments from the different contact
regions touch each other. The ``interaction'' between the filaments gives rise to the spreading resistance.
Because of the fractal nature of most surfaces the interaction between the heat flow 
filaments occur on many different length scales.}
\label{heatflowpic}
\end{figure}

\vskip 0.5cm
{\bf 6 Role of surface roughness}

Surface roughness has usually a strong influence on the heat transfer between 
macroscopic solids\cite{PLV}.
For hard solids the area of real contact may be very small compared to the nominal contact area, and
in these cases most of the heat may flow in the air film separating the non-contact region. The heat
transfer via the area of real contact is determined not just by the heat transfer resistance across the 
contacting interface (of atomic scale thickness) as studied above, 
but often most of the heat flow resistance is caused by the
so called spreading resistance, related to the interaction between the heat flow filaments which emerge 
from the areas of real contact. This latter contribution depends on the wide (fractal-like) distribution
of surface roughness length scales exhibited by most surfaces of macroscopic solids, see Fig. \ref{heatflowpic}. 
One can show that the 
total heat transfer resistance is (approximately) the sum of the two mentioned contributions:
$${1\over \alpha} \approx {1\over \alpha_{\rm spred}} + {1\over \alpha_{\rm c}}$$
where $1/\alpha_{\rm spred}$ is the spreading resistance studied in Ref. \cite{PLV}, and $1/\alpha_{\rm c}$
the resistance which determines the temperature jump (on atomistic length scale) 
across the area of real contact. One can show that (see Appendix F):
$${1\over \alpha_{\rm c}} \approx {1\over \alpha_{\rm b}} {1\over A_0 J_0^2}\int d^2 x \ J^2_z({\bf x}) $$
where $J_z({\bf x})$ is the heat current at the interface, $J_0$ the average heat current,  
and $\alpha_{\rm b}$ the (boundary) heat transfer coefficient
studied above (see Sec. 2). If the heat current would be constant through the area of real contact, then
$J_z = (A_0/A) J_0$, where $A$ is the area of real contact. In this case we get $\alpha_{\rm c} \approx (A/A_0) 
\alpha_{\rm b}$ and
$${1\over \alpha} \approx {1\over \alpha_{\rm spred}} + {1\over \alpha_{\rm b}} {A_0\over A}\eqno(25)$$
For most hard macroscopic solids the local pressure in the contact regions is very high,
which may result in ``cold welded'' contact regions with good thermal contact, in which case  
the contribution from the spreading resistance dominates the 
contact resistance. However, for weakly coupled microscopic solids the contribution from the
second term in (25) may be very important.

In Ref. \cite{Freitag} the temperature profile in graphene under current was 
studied experimentally. The heat transfer coefficient between graphene 
and the {\rm a}-${\rm SiO_2}$ substrate was determined by modeling the heat flow using the standard
heat flow equation with the heat transfer coefficient as the only unknown quantity. The authors found that using
a constant (temperature independent) heat transfer coefficient  $\alpha \approx 2.5\times 10^7 \ {\rm W/m^2K}$
the calculated temperature profiles in graphene are in good agreement with experiment. This $\alpha $ is about 
10 times smaller than expected for perfectly flat surfaces (see Sec. 5(c)).
In \cite{PUeba} we have studied the heat transfer between graphene and a-${\rm SiO_2}$.
We assumed that because of surface roughness 
the graphene only makes partial contact with the  ${\rm SiO_2}$ substrate,
which will reduce the heat transfer coefficient as compared to the perfect contact case.
The analysis indicated that the spreading resistance contribution
in (25) may be very important, and could explain the magnitude of the observed heat contact resistance. However,
assuming that (due to the roughness) $A/A_0 \approx 0.1$, the second term in (25) becomes of the same order
of magnitude as the measured heat contact resistance. Thus, in this particular application it is not clear which
term in (25) dominates the heat resistance, and probably both terms are important.

\vskip 0.5cm
{\bf 7 Summary}

To summarize, we have studied the heat transfer between coupled systems with flat interface.
We have presented simple analytical results which can be used to estimate the heat transfer coefficient.
The interaction between the solids is characterized by a spring constant (per unit area) $K$.
The formalism developed is general and valid both for strongly interacting ($K\rightarrow \infty$) 
and weakly interacting ($K\rightarrow 0$) solids.
We have shown that at low enough temperatures, even a very weak interfacial 
interaction will appear strong, and the
heat transfer is then given by the limiting formula obtained as $K\rightarrow \infty$.
Earlier analytical theories of heat transfer\cite{Pohl} does not account for the strength of the interaction between the
solids, but correspond to the limiting case  $K\rightarrow \infty$. However, we have shown that at room temperature
(or higher temperatures) the heat transfer between weakly interacting solids may be 100 times (or more) slower than
between strongly interacting solids.

Detailed results was presented for the heat transfer between a membrane (graphene) and a 
semi-infinite solid (a-${\rm SiO_2}$). For this case
the energy transfer is dominated by energy exchange between a bending vibrational mode 
of the graphene, and a Rayleigh surface phonon mode of the substrate. 
This  model assumes implicitly that, due to lattice non-linearity (and defects), there exist phonon scattering 
processes which rapidly transfer energy to the bending mode
involved in the heat exchange with the substrate. 
This may require high temperature in order for multi-phonon
scattering processes to occur at high enough rate.
The calculated value of the heat transfer coefficient was found to be in good agreement 
with the value deduced from the experimental data.

\vskip 0.2cm
{\bf Acknowledgments:}
We thank P. Avouris, Ch. W\"oll, G. Benedek and the authors of Ref. \cite{Chen} for useful communication.
B.N.J.P. was supported by Invitation Fellowship Programs for Research in Japan from
Japan Society of Promotion of Science (JSPS).
This work, as part of the European Science Foundation EUROCORES Program FANAS, was supported from funds 
by the DFG and the EC Sixth Framework Program, under contract N ERAS-CT-2003-980409.
H.U. was supported  by the Grant-in-Aid for Scientific
Research B (No. 21310086) from JSPS. A.I. Volokitin was supported by Russian Foundation
for Basic Research (Grant N 10-02-00297-a).

\vskip 0.5cm
{\bf Appendix A}

Here we prove equation (7). We get
$$\langle |u_{\rm 0 f}({\bf q},\omega)|^2\rangle =
{1\over (2 \pi)^6} \int d^2x dt d^2x' dt' $$
$$ \times  \langle u_{\rm 0f}({\bf x},t) u_{\rm 0f } ({\bf x}',t') \rangle e^{i[{\bf q}\cdot ({\bf x}-{\bf x}') - \omega(t-t')]}$$
$$={1\over (2 \pi)^6} \int d^2x dt d^2x' dt' $$
$$\times \langle u_{\rm 0 f}({\bf x} -{\bf x}' ,t-t') u_{\rm 0 f} ({\bf 0},0) \rangle 
e^{i[{\bf q}\cdot ({\bf x}-{\bf x}') - \omega(t-t')]}$$
$$={1\over (2 \pi)^6} \int d^2x dt d^2x' dt' \langle u_{\rm 0 f}  ({\bf x},t) u_{\rm 0 f}({\bf 0},0) \rangle 
e^{i[{\bf q}\cdot {\bf x} - i\omega t]}$$
$$= {A_0 t_0 \over (2\pi )^3} C_{uu}({\bf q},\omega)$$

\vskip 0.5cm
{\bf Appendix B}

Here we present an alternative derivation of Eq. (8).
Assume that the two solids interact weakly. In this case
the energy transfer from solid {\bf 0} to solid {\bf 1} is given by (6) with $K\rightarrow 0$:
$$\Delta E = (2\pi )^3 \int d^2 q d \omega
 \ \omega  K^2 {\rm Im} M_1({\bf q},\omega) 
\langle |u_{\rm 0 f}({\bf q},\omega)|^2\rangle. \eqno(B1)  $$
At thermal equilibrium this must equal the energy transfer from solid {\bf 1}
to solid {\bf 0} given by
$$\Delta E = (2\pi )^3 \int d^2 q d \omega
 \ \omega  K^2 {\rm Im} M_0({\bf q},\omega) 
\langle |u_{\rm 1 f}({\bf q},\omega)|^2\rangle. \eqno(B2)  $$
From (B1) and (B2) we get
$${\rm Im} M_1({\bf q},\omega) 
\langle |u_{\rm 0 f}({\bf q},\omega)|^2\rangle 
={\rm Im} M_0({\bf q},\omega) 
\langle |u_{\rm 1 f}({\bf q},\omega)|^2\rangle \eqno(B3) $$ 
We now assume that solid {\bf 1} is a layer of non-interacting harmonic oscillators.
Thus if  $\rho_1$ is the mass per unit area we have
$$\rho_1 \ddot u_1 +\omega_1^2 u_1 = \sigma $$
or
$$u_1({\bf q},\omega) = {\sigma({\bf q},\omega) \over \rho_1(\omega_1^2-\omega^2)-i0^+}$$
Thus
$$M_1({\bf q},\omega) = {1\over \rho_1(\omega_1^2-\omega^2)-i0^+}$$  
and for $\omega > 0$
$${\rm Im} M_1({\bf q},\omega) = {\pi \over 2 \rho_1 \omega_1}\delta (\omega-\omega_1)\eqno(B4)$$
We write $u_1$ in the standard form:
$$u_1 = {1\over (2 \pi )^2}\int d^2q \left ({\hbar \over 2 \rho_1 \omega_1}\right )^{1/2}$$ 
$$ \times \left (b_{\bf q}e^{i({\bf q}\cdot {\bf x} - \omega_1 t)}
+b^+_{\bf q}e^{-i({\bf q}\cdot {\bf x} - \omega_1 t)}\right ),$$
so that for $\omega >0$:
$$u_1({\bf q},\omega) = {1\over (2 \pi )^2}\left ({\hbar \over 2 \rho_1 \omega_1}\right )^{1/2} b_{\bf q} \delta (\omega-\omega_1)$$ 
Thus we get
$$\langle |u_1({\bf q},\omega)|^2 \rangle = {t_0\over (2 \pi )^5}{\hbar \over 2 \rho_1 \omega_1}
{1\over 2} \langle b_{\bf q}b^+_{\bf q} +b^+_{\bf q} b_{\bf q} \rangle \delta (\omega-\omega_1)$$ 
where we have used that
$$\left [ \delta (\omega-\omega_1) \right ]^2 = \delta (\omega-\omega_1) {1\over 2 \pi}\int dt = \delta (\omega-\omega_1) {t_0\over 2 \pi}$$
Using that
$$\langle b_{\bf q}b^+_{\bf q} +b^+_{\bf q} b_{\bf q} \rangle = [2 n(\omega_1)+1](2\pi)^2 \delta({\bf q}-{\bf q}) = [2 n(\omega_1)+1] A_0$$ 
we get 
$$ \langle |u_{\rm 1 f}({\bf q},\omega)|^2\rangle = {A_0 t_0 \over (2 \pi )^5} {\hbar \over 2 \rho_1 \omega_1} \left ( n(\omega)+{1\over 2}\right )
\delta (\omega-\omega_1)\eqno(B5)$$
Combining (B3)-(B5) gives
$$\langle |u_{\rm 0 f}({\bf q},\omega)|^2\rangle =  {2 A_0 t_0 \hbar \over (2 \pi )^6}  
\left ( n(\omega)+{1\over 2}\right ) {\rm Im} M_0({\bf q},\omega)\eqno(B6)$$  

\vskip 0.5cm
{\bf Appendix C}

Eq. (8) is a standard result 
but the derivation is repeated here for the readers
convenience. Let us write the Hamiltonian as
$$H=H_0+\int d^2x \ u({\bf x},t) \sigma({\bf x},t) $$ 
where $\sigma({\bf x},t)$ is an external stress acting on the surface $z=0$ of the solid.
We first derive a formal expression for $ M ({\bf q},\omega)$ defined by the linear response formula
$$\langle u({\bf q},\omega) \rangle = M ({\bf q},\omega) \sigma({\bf q},\omega)$$
We write $\langle u \rangle = {\rm Tr}(\rho u)$ where the density operator satisfies
$$i\hbar {\partial \rho \over \partial t} = [H,\rho] $$
We write $\rho=\rho_0+\rho_1$ and get
$$\rho_1 = {1\over i \hbar} \int_{-\infty}^t dt' e^{-iH_0 (t-t')/\hbar} [V(t'),\rho_0] e^{iH_0 (t-t')/\hbar}$$
Thus using $\langle u \rangle = {\rm Tr} (\rho_1 u) $ we get 
$$\langle u \rangle = {1\over i \hbar} \int d^2 x' dt' \ \theta (t-t') \langle [u({\bf x},t),u({\bf x}',t')] \rangle \sigma({\bf x}',t')$$
$$= {1\over i \hbar} \int d^2 x' dt' \theta(t-t') \langle [u({\bf x}-{\bf x}',t-t'),u({\bf 0},0)]\rangle  \sigma({\bf x}',t')$$
Thus
$$M({\bf q},\omega)= 
{1\over i \hbar} \int d^2x  dt \ \theta(t) \langle  [u({\bf x},t),u({\bf 0},0)]\rangle e^{-i({\bf q}\cdot {\bf x}-\omega t)}\eqno(C1)$$
where 
$$u({\bf x},t) = e^{-iH_0 t/\hbar} u({\bf x},0)e^{iH_0 t/\hbar}.$$

Let $|n\rangle $ be an eigenstate of $H_0$ corresponding to the energy $E_n$. Using (C1) we get
$$M({\bf q},\omega)= {1\over i \hbar}  \int d^2 x dt \ \theta (t) \sum_{nm} Z^{-1} e^{-\beta E_n} e^{-i({\bf q}\cdot {\bf x}-\omega t)}$$
$$\times \big ( \langle n| u({\bf x},0)|m \rangle \langle m| u({\bf 0},0)|n \rangle e^{-i(E_n-E_m)t/\hbar} $$
$$-\langle n| u({\bf 0},0)|m \rangle \langle m| u({\bf x},0)|n \rangle e^{i(E_n-E_m)t/\hbar} \big ) $$
$$= {1\over i \hbar}  \int d^2 x \sum_{nm} Z^{-1} e^{-\beta E_n} e^{-i{\bf q}\cdot {\bf x}}$$
$$\times \bigg ( {\langle n| u({\bf x},0)|m \rangle \langle m| u({\bf 0},0)|n \rangle \over i(E_n-E_m)/\hbar -i\omega+0^+} $$
$$+{\langle n| u({\bf 0},0)|m \rangle \langle m| u({\bf x},0)|n \rangle \over i(E_n-E_m)/\hbar +i\omega-0^+} \bigg )\eqno(C2) $$
where $\beta = 1/k_{\rm B} T$ and $Z=\sum_n {\rm \exp}(-\beta E_n)$. From (C2) we get
$${\rm Im} M({\bf q},\omega)= {\pi \over \hbar}  \int d^2 x \sum_{nm} Z^{-1} e^{-\beta E_n} e^{-i{\bf q}\cdot {\bf x}}$$
$$\times \big ( \langle n| u({\bf x},0)|m \rangle \langle m| u({\bf 0},0)|n \rangle (-\delta [\omega -  (E_n-E_m)/\hbar]) $$
$$+\langle n| u({\bf 0},0)|m \rangle \langle m| u({\bf x},0)|n \rangle \delta [\omega + (E_n-E_m)/\hbar ]\big )\eqno(C3) $$
Changing summation index from $(n,m)\rightarrow (m,n)$ the $(m,n)$-dependent part of the second term in (C3) can be rewritten as
$$\sum_{nm} e^{-\beta E_m}  \langle m| u({\bf 0},0)|n \rangle \langle n| u({\bf x},0)|m \rangle $$
$$\times \delta [\omega - (E_n-E_m)/\hbar] $$
$$= \sum_{nm} e^{-\beta E_n} e^{-\beta (E_m-E_n)}  \langle n| u({\bf x},0)|m \rangle   \langle m| u({\bf 0},0)|n \rangle $$
$$\times \delta [\omega - (E_n-E_m)/\hbar] $$
$$=   e^{\beta \omega} \sum_{nm} e^{-\beta E_n} \langle n| u({\bf x},0)|m \rangle   \langle m| u({\bf 0},0)|n \rangle $$
$$\times \delta [\omega - (E_n-E_m)/\hbar] $$
Replacing the second term in (C3) with this expression gives
$${\rm Im} M({\bf q},\omega)={1\over 2\hbar}\left (e^{\beta \hbar \omega}-1\right )$$
$$\times \int d^2x dt \ e^{-i({\bf q}\cdot {\bf x} -i\omega t)}
\langle u({\bf x},t)u({\bf 0},0)\rangle $$
$$={1\over 2\hbar}\left (e^{\beta \hbar \omega}-1\right ) (2\pi)^3 C_{uu}({\bf q},\omega)$$
From the last equation follows the fluctuation-dissipation theorem:
$$C_{uu}({\bf q},\omega) = {1 \over (2\pi)^3} { 2\hbar \over e^{\beta \hbar \omega}-1} {\rm Im} M({\bf q},\omega)  $$

\vskip 0.5cm
{\bf Appendix D}

We assume high temperatures and interfacial disorder. In this case the elastic waves generated by the 
stochastic pulsating forces between the
atoms at the interface give rise to (nearly) incoherent emission of sound waves (or phonons). Thus, we can obtain the total
energy transfer by just adding up the contributions from the elastic waves emitted from each interfacial
atom. Assume for simplicity that the interfacial atoms of solid {\bf 0} for a simple square lattice with lattice constant
$a_0$. Consider the atom at ${\bf x = 0}$ and let $u_0(t)$ denote the vertical displacement of the atom. The 
force
$$F(t)=k[u_0(t)-u_1(t)],$$
or
$$F(\omega )=k[u_0(\omega )-u_1(\omega )],\eqno(D1)$$
is acting on solid {\bf 1} at ${\bf x = 0}$. We can write $k=Ka_0^2$ where $K$ is the force constant per unit area (see Sec. 2).
The force $F(t)$ gives a stress
$$\sigma ({\bf x},t) = F(t) \delta ({\bf x})$$
acting on solid {\bf 1}. We can also write 
$$\sigma ({\bf q},\omega) =  (2\pi)^{-2}  F(\omega)  $$
Note that
$$u_1(\omega ) = u_1({\bf x = 0},\omega) = \int d^2q \ u_1({\bf q},\omega) $$
$$= \int d^2q \ M_1 ({\bf q},\omega) \sigma ({\bf q},\omega)$$
$$ = {1\over (2\pi)^2} \int d^2q \ M_1 ({\bf q},\omega) F(\omega)$$
$$=\bar M_1(\omega) F(\omega), \eqno(D2)$$
where 
$$\bar M_1 (\omega) = {1\over (2\pi)^2} \int d^2q \ M_1 ({\bf q},\omega)$$
In a similar way one get
$$u_0 (\omega ) = u_{\rm 0 f} (\omega)- \bar M_0(\omega) F(\omega). \eqno(D3)$$ 
Combining (D1)-(D3) gives
$$u_1 (\omega) = {k \bar M_1(\omega) \over 1+k [\bar M_0(\omega)+ \bar M_1(\omega)]} u_{\rm 0 f} (\omega), \eqno(D4)$$
$$u_0 (\omega) = {1+k \bar M_1(\omega) \over 1+k [\bar M_0(\omega)+ \bar M_1(\omega)]} u_{\rm 0 f} (\omega). \eqno(D5)$$
The energy transferred to solid ${\bf 1}$ from solid ${\bf 0}$ during the time period $t_0$ can be written as
$$\Delta E = N \int d t \  \dot u_1(t) F (t),$$
where  $N=A_0/a_0^2$ is the number of interfacial atoms of solid {\bf 0}. One can also write 
$$\Delta E = 2\pi N \int  d \omega \  (-i\omega) u_1(\omega) F (-\omega)$$
Using (D1) and (D4) and (D5) we obtain
$$\Delta E = 2\pi  N \int d \omega {\omega  k^2 {\rm Im} \bar M_1(\omega) \over |1+k [\bar M_0(\omega)+ \bar M_1(\omega)]|^2} 
\langle |u_{\rm 0 f}(\omega)|^2\rangle, \eqno(D6)  $$
where we have performed an ensemble (or thermal) average denoted by  $ \langle .. \rangle $. 
Next, note that
$$\langle |u_{\rm 0 f}(\omega)|^2\rangle =
{1\over (2 \pi)^2} \int dt dt' \ \langle u_{\rm 0f}(t) u_{\rm 0f } (t') \rangle e^{-i \omega(t-t')}$$
$$={1\over (2 \pi)^2} \int dt dt' \langle u_{\rm 0 f}(t-t') u_{\rm 0 f} (0) \rangle 
e^{-i\omega (t-t')}$$
$$={1\over (2 \pi)^2} \int dt dt' \langle u_{\rm 0 f}  (t) u_{\rm 0 f}(0) \rangle 
e^{i \omega t} = {2t_0 \over 2\pi} \bar C_{uu}(\omega), \eqno(D7)$$
where 
$$\bar C_{uu}(\omega) =  {1\over 2\pi } \int dt \ 
\langle u_{\rm 0 f} (t) u_{\rm 0 f} (0)\rangle e^{-i \omega t}, $$
is the displacement correlation function.
Note that
$$\bar C_{uu}(\omega) =   \int d^2q  \  C_{uu}({\bf q},\omega) $$
Thus, using (8) we get
$$\bar C_{uu}(\omega)= {2\over (2\pi)^3} {\Pi (\omega) \over \omega} {\rm Im}  \int d^2q \ M_0 ({\bf q}, \omega) $$
$$= {2\over 2\pi} {\Pi (\omega) \over \omega}  {\rm Im} \bar M_0 (\omega)\eqno(D8)$$
Substituting (D7) in (D6) and using (D8) gives the heat current $J_{0\rightarrow 1} = \Delta E/ A_0 t_0$ from solid  ${\bf 0}$ to  solid ${\bf 1}$:
$$J_{0\rightarrow 1} 
= {4 A^* \over (2 \pi)^3} \int_0^\infty d \omega \ \Pi (\omega)
{ {\rm Im} K M_0(\omega ) {\rm Im} K M_1(\omega )
\over | 1+K[M_0(\omega)+M_1(\omega)]|^2}, $$
where $ A^* = (2\pi )^2/ a_0^2$ is the area of the Brillouin zone and
where we have defined
$$M(\omega) = a_0^2 \bar M(\omega) = {1\over A^*} \int_{q<q_{\rm c}} d^2q \ M({\bf q},\omega)$$
where the $q$-integral is over $|{\bf q}| < q_{\rm c}$ with $\pi q_{\rm c}^2 = A^*$.
A similar equation with $T_0$ replaced by $T_1$ gives the energy transfer from solid ${\bf 1}$ to solid ${\bf 0}$, and the
net energy flow $J=J_{0\rightarrow 1}-J_{1\rightarrow 0}$.
The heat transfer coefficient $\alpha = (J_{0\rightarrow 1}-J_{1\rightarrow 0})/(T_0-T_1)$ gives in the limit
$(T_0-T_1)\rightarrow 0$:
$$\alpha 
= {4 A^* \over (2 \pi)^3} \int_0^\infty d \omega \ {\partial \Pi (\omega)\over \partial T} 
{ {\rm Im} K M_0(\omega ) {\rm Im} K M_1(\omega )
\over | 1+K[ M_0(\omega)+M_1(\omega)]|^2}. 
\eqno(D9)$$
The derivation above is only valid for high temperature where $k_{\rm B}T > \hbar \omega_0$, where $ \hbar \omega_0$ 
is the highest phonon energy of solid {\bf 0}. 
However, we can apply the theory (approximately) to all temperatures if we take 
the cut-off wavevector  $q_{\rm c} $ to be the smallest of  $q_1$ and $q_2$, where
$\pi q_1^2 = (2\pi )^2/a_0^2$ (where  $a_0 $ is the lattice 
constant) and where $q_2=k_{\rm B}T/\hbar c_0 $ (where  $c_0 $ is the smallest sound velocity of solid {\bf 0}) is
the thermal wavevector.

\vskip 0.5cm
{\bf Appendix E}

Here we show that (20) reduces to (19) for high temperatures and when solid {\bf 0} is described by an Einstein model.
At high temperatures and weak interfacial coupling, (20) becomes
$$\alpha 
= {4 k_{\rm B} A^* \over (2 \pi)^3} \int_0^\infty d \omega \ {\rm Im} K M_0(\omega ) {\rm Im} K M_1(\omega )\eqno(E1)$$
We assume for solid {\bf 0} that
$$M_0(\omega) = {1 \over \rho_0 (\omega_0^2-\omega^2)-i0^+}$$
where $\rho_0 = m_0/a_0^2$ is the mass per unit area.
Thus 
$${\rm Im} M_0(\omega) = {\pi \over 2 \rho_0\omega_0}\delta (\omega_0-\omega)$$
Substituting this in (E1) gives
$$\alpha 
= {4 k_{\rm B} K^2 A^* \over (2 \pi)^3} {\pi \over 2 \rho_0\omega_0}  {\rm Im} M_1(\omega_0 )$$
$$= { k_{\rm B} K^2 \over (2 \pi)^2} {1 \over \rho_0 \omega_0}  \int d^2q \ {\rm Im} M_1({\bf q},\omega_0 )$$
Substituting () into this equation and denoting $q=(\omega_0 /c_{\rm T})x^{1/2}$ gives
$$\alpha = { k_{\rm B} K^2 \over \rho_0 \rho_1 c_{\rm T}^3}$$
$$ \times  {1\over 8 \pi} {\rm Re} \int_0^\infty 
dx  {2 \left (\gamma-x\right )^{1/2} 
\over \left (1-2x\right  )^2 +4 \left (1-x\right )^{1/2} \left (\gamma-x\right )^{1/2}} \eqno(E2)$$
where $\gamma = (c_{\rm T}/c_{\rm L})^2$, which agree with (19).

\vskip 0.5cm
{\bf Appendix F}

Let $1/\alpha_{\rm c}$ be the interfacial contact resistance associated with
the jump in the temperature (on an atomistic scale) in the contact area between two solids, 
and let $1/\alpha_{\rm spred}$ be the spreading resistance
associated with the interaction between the heat filaments emerging from all the contact regions.
Since these two resistances act in
series one expect the total contact resistance to be the sum of the two contributions, i.e., 
$${1\over \alpha} \approx {1\over \alpha_{\rm spred}} + {1\over \alpha_{\rm c}} $$ 
We can prove this equation and Eq. (24) using the formalism developed in Ref. \cite{PLV}. We assume that all the heat energy
flow via the area of real contact. In this case the interfacial heat current $J_z({\bf x})$ vanish in the non-contact area. In the area
of real contact the temperature $T({\bf x},z)$ change abruptly (on an atomistic scale) when
$z$ increases from $z=-0^+$ (in solid {\bf 0}) to $z=0^+$ (in solid {\bf 1}), and the jump determines the heat current:
$J_z({\bf x}) = \alpha_{\rm b} [T({\bf x},-0)-T({\bf x},+0)]$. If we denote $\psi ({\bf x}) = T({\bf x},-0)-T({\bf x},+0)$ the
equation
$$J_z({\bf x})[J_z({\bf x})-\alpha_{\rm b}\psi ({\bf x})] = 0$$
will be valid everywhere at the interface. From this equation we get
$$\int d^2q' J_z ({\bf q}-{\bf q}') [J_z({\bf q}')-\alpha_{\rm b}\psi ({\bf q}')] = 0$$
Following the derivation in Sec. 2.2.1 in Ref. \cite{PLV} we get instead of Eq. (20) in  Ref. \cite{PLV} the equation
$${1\over \alpha} = {(2\pi )^2\over \kappa} {1\over A_0 J_0^2} \int d^2q \ 
{1\over q} \langle |\Delta J_z({\bf q})|^2\rangle $$
$$+ {1\over  \alpha_{\rm b}} {1\over A_0J_0^2} \int d^2x J_z^2 ({\bf x})
= {1\over \alpha_{\rm spred}} + {1\over \alpha_{\rm c}}\eqno(F1)$$
Here $J_0$ is the average or nominal heat current, $\Delta J_z({\bf x}) =J_z({\bf x}) -J_0$, 
and $\kappa $ an effective heat conductivity ($\kappa^{-1} = \kappa_0^{-1}+ \kappa_1^{-1}$). 
The first term in (F1) is the spreading resistance studied in Ref. \cite{PLV} while the second term
is the contribution from the temperature jump on the atomistic scale accross the area of real contact.


\begin{thebibliography}{99}
\bibitem{explain}
In Ref. \cite{Ryberg} we have calculated $\xi' = \xi_\perp /8\pi$ within the elastic continuum model. It only depends on the ratio
$\lambda = c_{\rm T}/c_{\rm L}$. For $\lambda = 0.5$ the contribution to $\xi_\perp$ from emission of surface phonons is 2.10
and the contribution from  bulk longitudinal and transverse phonons is 1.19, giving  $\xi' = (2.10+1.19) /8\pi \approx 0.13$.
\bibitem{addpara}
In the calculation we have used:
Sound velocity and mass density of  $^4$He: $c_{L0} =  238 \ {\rm m/s}$
and $\rho_0 =  150 \ {\rm kg/m^3}$. 
Sound velocity and mass density of a-${\rm SiO}_2$:
$c_{T1} =  3743 \ {\rm m/s}$,
$c_{L1} =  5953 \ {\rm m/s}$,
and $\rho_1 =  2200 \ {\rm kg/m^3}$.
\end{thebibliography}
\end{document}